\documentstyle[12pt]{article}

\catcode`@=11

\def\nopagenumbers{\pagestyle{empty}}              
    \nopagenumbers   

\def\pagesetup{
\textwidth 6.0in
\textheight 8.5in
\pagestyle{empty}
\topmargin -0.25truein
\oddsidemargin 0.30truein
\evensidemargin 0.30truein
\raggedbottom\parindent=20pt
\baselineskip=14pt}

\pagesetup

\def\@normalsize{\@setsize\normalsize{14pt}\xiipt\@xiipt
\abovedisplayskip 12pt plus3pt minus7pt%
\belowdisplayskip \abovedisplayskip
\abovedisplayshortskip  \z@ plus3pt%
\belowdisplayshortskip  6.5pt plus3.5pt minus3pt%
\let\@listi\@listI}   

\def\small{\@setsize\small{12pt}\xpt\@xpt
\abovedisplayskip 10pt plus2pt minus5pt%
\belowdisplayskip \abovedisplayskip
\abovedisplayshortskip  \z@ plus3pt%
\belowdisplayshortskip  6pt plus3pt minus3pt
\def\@listi{\leftmargin\leftmargini 
\topsep 6pt plus 2pt minus 2pt\parsep 3pt plus 2pt minus 1pt
\itemsep \parsep}}

\def\footnotesize{\@setsize\footnotesize{11pt}\ixpt\@ixpt
\abovedisplayskip 8.5pt plus 3pt minus 4pt%
\belowdisplayskip \abovedisplayskip
\abovedisplayshortskip \z@ plus2pt%
\belowdisplayshortskip 4pt plus2pt minus 2pt
\def\@listi{\leftmargin\leftmargini 
\topsep 4pt plus 2pt minus 2pt\parsep 2pt plus 1pt minus 1pt
\itemsep \parsep}}

\newcommand{\symbolfootnotes}{\renewcommand{\thefootnote}
	{\fnsymbol{footnote}}}

\newcommand{\alphafootnotes}
	{\setcounter{footnote}{0}
	 \renewcommand{\thefootnote}{\alph{footnote}}}

  \alphafootnotes
\newenvironment{Titlepage}{\parsep=0pt \topsep=0pt \symbolfootnotes}%
             {\relax}

\def\Title#1\endTitle{\begin{center}%
   \baselineskip=16pt 
 \bf #1\\[.5cm]}                    
\def\Author#1#2\endAuthor{\small\it
   {\rm #1}\\[1pt] #2\\[.3cm]}
\def\endAuthors{\end{center}
                \vglue .5cm    
                      \alphafootnotes}

\newenvironment{Abstract}%
       {\centering\bgroup
          \begin{minipage}{30pc}\small
            \noindent
	    \centerline{\tenrm ABSTRACT}
	    \parindent=0pt}%
         {\end{minipage}\egroup\par
         \normalsize}

\chardef\bslash=`\\    

\catcode`@=12   

\def\be{\begin{equation}} \def\ee{\end{equation}}

\catcode`@=11

\def\eqalign#1{\null\,\vcenter{\openup\jot\m@th
  \ialign{\strut\hfil$\displaystyle{##}$&$\displaystyle{{}##}$\hfil
      \crcr#1\crcr}}\,}
\def\meqalign#1{\null\,\vcenter{\openup\jot\m@th
  \ialign{\strut\hfil$\displaystyle{##}$&&$\displaystyle{{}##}$\hfil
      \crcr#1\crcr}}\,}
\catcode`@=12   

\def\umlaut{\"} 

\def\journalfont{\rm}         
\def\jou#1{{\journalfont #1\ }}
\def\joudef#1#2{\def #1{\jou{\ignorespaces #2}}}

\joudef{\aaa}    { Astron.\ Astrophys.}
\joudef{\aip}    { Adv.\ Phys.}
\joudef{\adm}    { Adv.\ Math.}
\joudef{\am}     { Ann.\ Math.}
\joudef{\apny}   { Ann.\ Phys.\ (N.Y.)}
\joudef{\apj}    { Astrophys.\ J.}
\joudef{\cjp}    { Can.\ J.\ Phys.}
\joudef{\cmp}    { Commun.\ Math.\ Phys.}
\joudef{\cqg}    { Class.\ Quantum Grav.}
\joudef{\grg}    { Gen.\ Rel.\ Grav.}
\joudef{\ijmpd}  { Int.\ J.\ Mod.\ Phys.\ D}
\joudef{\ijtp}   { Int.\ J.\ Theor.\ Phys.}
\joudef{\im}     { Invent.\ Math.}
\joudef{\jm}     { J.\ Math.}
\joudef{\jmp}    { J.\ Math.\ Phys.}
\joudef{\jpa}    { J.\ Phys.\ A}
\joudef{\mnras}  { Mon.\ Not.\ R.\ Ast.\ Soc.}
\joudef{\mpla}   { Mod.\ Phys.\ Lett.\ A}
\joudef{\nature} { Nature}
\joudef{\nc}     { Nuovo Cim.}
\joudef{\npb}    { Nuc.\ Phys.\ B}
\joudef{\pla}    { Phys.\ Lett. A}
\joudef{\plb}    { Phys.\ Lett. B}
\joudef{\pr}     { Phys.\ Rev.}
\joudef{\prd}    { Phys.\ Rev.\ D}
\joudef{\prep}   { Phys.\ Rep.}
\joudef{\prl}    { Phys.\ Rev.\ Lett.}
\joudef{\prsla}  { Proc.\ Roy.\ Soc.\ Lond.\ A}
\joudef{\ptp}    { Prog.\ Theor.\ Phys.}
\joudef\rmp      { Rev.\ Mod.\ Phys.}
\joudef\spj      { Sov.\ Phys.\ JETP}

\def\Hscr{{\cal H}} \def\Nscr{{\cal N}}
\def\bfA{{\bf A}}  \def\bfL{{\bf L}}
 
\def\dbL{\dot{\:\bfL}}

\def\Tr{\mathop{\rm Tr}\nolimits}

\def\fraction#1#2{{\textstyle{#1\over#2}}}

\begin{document} \bibliographystyle{prsty}

\begin{Titlepage}

\Title{A tensorial Lax pair equation and integrable systems \\
       in relativity and classical mechanics
 \footnote{to appear in the proceedings of the 7th Marcel Grossman Meeting on
           General Relativity}
 \footnote{gr-qc/9410011}}
\endTitle
\Author{Kjell Rosquist}
RGGR group, Chimie-Physique, CP231, ULB Campus Plaine, Brussels, Belgium \\
and \\
Department of Physics, Stockholm University, Box 6730, 113 85 Stockholm,
  Sweden
\endAuthor
\endAuthors

\begin{Abstract}
It is shown that the Lax pair equation $\dbL = [\bfL,\bfA]$ can be given a
neat tensorial interpretation for finite-dimensional quadratic Hamiltonians.
The Lax matrices $\bfL$ and $\bfA$ are shown to arise from third rank tensors
on the configuration space. The second Lax matrix $\bfA$ is related to a
connection which characterizes the Hamiltonian system. The Toda lattice
system is used to motivate the definition of the Lax pair tensors. The
possible existence of solutions to the Einstein equations having the Lax
pair property is discussed.
\end{Abstract}

\end{Titlepage}

\section{Introduction}

A characteristic feature in the study of integrable systems is the existence
of a pair of matrices (Lax pair), $\bfL$, $\bfA$, satisfying the Lax equation
\cite{lax:pair}
\be\label{eq:lax}
    \dbL = [\bfL,\bfA] \ .
\ee
It follows from the Lax equation that the quantities $I_k = k^{-1}\Tr\bfL^k$
are constants of the motion. If sufficiently many of the $I_k$ are
independent and in involution ({\it i.e.\/} $[I_k, I_\ell] =0$ for all $k$,
$\ell$) then the system is said to be integrable in the Liouville sense.
In this letter we give a new characterization of the Lax pair property in
terms of a tensorial equation valid for any finite-dimensional quadratic
Hamiltonian system. In the new formulation the matrix $\bfA$ is related to a
certain dynamical connection. Geometric interpretations of the Lax pair
property have appeared before \cite{perelomov:int,fmv:geomlax}. However, the
approach proposed in this contribution is more direct in the sense that it
relates the Lax pair property to configuration space tensors rather than
phase space tensors.

We shall consider finite-dimensional Hamiltonians with a quadratic kinetic
energy, $H = \frac12 h^{\alpha\beta} p_\alpha p_\beta + V(q)$. In typical
problems in classical mechanics, the kinetic metric is positive definite and
constant. In general $h^{\alpha\beta}$ has an arbitrary signature and
depends on the configuration variables $q^\alpha$. Our approach is to
transform the system to a geometric form whereby the orbits will correspond to
geodesics in a certain dynamical geometry. We look for a way to write the
Lax pair equation in a covariant form with respect to that dynamical
geometry. As a guide in this search we shall use the three particle Toda
lattice with free ends. Its Hamiltonian is given by \cite{perelomov:int}
\be\eqalign{
      H &= \fraction12 \left( p_1{}^2 + p_2{}^2 + p_3{}^2 \right)
                                                  + V_3 + V_1 \ ,\cr
    V_1 &= a_1{}^2 \ ,\qquad V_3 = a_3{}^2 \ ,\cr
    a_1 &= e^{q^2-q^3} \ ,\qquad a_3 = e^{q^1-q^2} \ .\cr
}\ee
The system has a Lax matrix formulation $\dbL \equiv \{\bfL, H\} =
[\bfL,\bfA]$ where
\be
    \bfL = \left(
           {\matrix{{p_1}&{a_3}&0\cr{a_3}&{p_2}&{a_1}\cr0&{a_1}&{p_3}\cr }}
           \right) \ , \qquad
    \bfA = \left( {\matrix{0&{a_3}&0\cr{-a_3}&0&{a_1}\cr0&{-a_1}&0\cr }}
           \right) \ .
\ee
The Lax matrix gives rise to the three invariants
\be\label{eq:inv}\eqalign{
     I_1 &= \Tr\bfL = p_1 + p_2 + p_3 \ ,\cr
     I_2 &= \fraction12 \Tr\bfL^2 = H \ ,\cr
     I_3 &= \fraction13 \Tr\bfL^3
                         = I_1 I_2 - \fraction16 I_1{}^3 + \hat I_3 \ ,\cr
}\ee
where
\be
     \hat I_3 = p_1 p_2 p_3 - V_1 p_1 - V_3 p_3 \ .
\ee
These invariants are in involution and the system is therefore Liouville
integrable. We mention that the invariants $I_k$ can be interpreted as
coadjoint invariants of the solvable Lie group of real upper triangular
matrices with unit determinant \cite{kostant:toda,adler:kdv,symes:toda}. In
order to obtain a tensorial formulation of the Lax pair equation we
formulate the Toda system geometrically in terms of a certain dynamical
metric.

\section{The dynamical geometry}
We now discuss how the dynamical metric may be defined. For any given
Hamiltonian system which is not explicitly time dependent ($\partial
H/\partial t = 0$) let us reparametrize the time variable
according to the recipe $dt \rightarrow \Nscr^{-1} dt$, $H \rightarrow
\Hscr_\Nscr = \Nscr H$ \cite{lanczos:mechanics,urj:geom}, where $\Nscr$
is some function on the phase space. Hamilton's equations with respect to
the new Hamiltonian are equivalent to the original equations of motion at zero
energy ($H=0$). To transform the dynamics for a nonzero value of the energy
one must first subtract that value from the original Hamiltonian before
making the time reparametrization. Suppose now that we are dealing with a
Hamiltonian of the type $H = \frac12 h^{\alpha\beta}p_\alpha p_\beta + V(q)$.
The special choice $\Nscr = [2(E-V)]^{-1}$ leads to the Jacobi time gauge in
which the potential energy of the reparametrized Hamiltonian, $\Hscr_\Nscr$,
is just a constant. Subtracting that constant we obtain the Jacobi Hamiltonian
$H_J = \frac14 (E-V)^{-1} h^{\alpha\beta} p_\alpha p_\beta$. The dynamics is
now equivalent to geodesic motion in the Jacobi geometry given by $ds^2 =
2(E-V  ) h_{\alpha\beta} dq^\alpha dq^\beta$. This procedure was used by
Rosquist and Pucacco \cite{rp:inv} to give a unified derivation of invariants
at both fixed and arbitrary energy generalizing Darboux's century old
condition for 2-dimensional Hamiltonians admitting a second quadratic
invariant. However, the method does suffer from a drawback in that the
resulting Jacobi geometry is energy dependent. That is to say that the full
dynamics is not represented by a single geometry but rather corresponds to a
family of geometries parametrized by the energy. One way of avoiding this
drawback is to use a suitable canonical transformation to transform the
original Hamiltonian to a geometric form (see {\it e.g.\/} Ashtekar {\it et
al.\/} \cite{atu:miniquant}). However, such a transformation is not always
possible to find and in general it destroys the conformal flatness of the
original problem. We shall now give another method to transform a Hamiltonian
to a geometric form which represents the dynamics in terms of a single
geometry.

The transformation we are about to describe is a variation of the time
reparametrization scheme. In fact, we shall work in an extended
Hamiltonian framework \cite{lanczos:mechanics}. The key point in this
approach is to introduce an additional configuration space variable
$q^{n+1}$ together with an associated canonical momentum $p_{n+1}$. In this
extended phase space the dynamics is given by a constrained Hamiltonian
$\Hscr(q_1,\ldots,q^{n+1},p_1,\ldots,p_{n+1}) = 0$. The constraint is usually
chosen as $\Hscr=H-p_{n+1}$ where $p_{n+1} = E$. This gives an extended
Hamiltonian which is no longer quadratic in the momentum variables. However,
as pointed out by Lanczos \cite{lanczos:mechanics}, other choices of extended
Hamiltonian are allowed. For a system at positive energy we exploit this
freedom to take $\Hscr=H-\frac12 p_{n+1}^2$ as our extended Hamiltonian. This
means that the additional momentum is related to the energy by $p_{n+1} =
\sqrt{2E}$.

\section{Geometrical formulation of the Toda lattice}
Before applying the above time reparametrization scheme to the Toda
Hamiltonian we first adapt the coordinate to the linear symmetry by
performing the orthogonal transformation
\be\label{eq:orthotrans}\meqalign{
    q^1 &=  \fraction1{\sqrt2}\bar q^1 + \fraction1{\sqrt6}\bar q^2
                             + \fraction1{\sqrt3}\bar q^3 \ ,\qquad\qquad
   &p_1 = \fraction1{\sqrt2}\bar p_1 + \fraction1{\sqrt6}\bar p_2
                                   + \fraction1{\sqrt3}\bar p_3 \ ,\cr
    q^2 &= -\sqrt{\fraction23}\,\bar q^2+\fraction1{\sqrt3}\bar q^3 \ ,
   &p_2 = -\sqrt{\fraction23}\,\bar p_2 + \fraction1{\sqrt3}\bar p_3 \ ,\cr
    q^3 &= -\fraction1{\sqrt2}\bar q^1 + \fraction1{\sqrt6}\bar q^2
                                   + \fraction1{\sqrt3}\bar q^3 \ ,
   &p_3 = -\fraction1{\sqrt2}\bar p_1 + \fraction1{\sqrt6}\bar p_2
                                   + \fraction1{\sqrt3}\bar p_3 \ ,\cr
}\ee
The Hamiltonian then becomes
\be
    H = \frac12 (\bar p_1{}^2 + \bar p_2{}^2 + \bar p_3{}^2)
        + 2e^{\sqrt2\bar q^1}
           \cosh\left(\sqrt6\,\bar q^2\right) \ .
\ee
The coordinate $\bar q^3$ is now cyclic with constant conjugate momentum
$\bar p_3$. The cubic invariant becomes
\be
     \hat I_3 = \sqrt6 \tilde I_3 + \fraction{5\sqrt3}{18} \bar p_3{}^3
                 - \fraction1{\sqrt3} \bar p_3 H \ ,
\ee
where
\be
     \tilde I_3 = - \fraction1{18} \bar p_2{}^3
         + \fraction16 \bar p_1{}^2 \bar p_2
         + \fraction1{\sqrt3} e^{\sqrt2\bar q^1}\sinh(\sqrt6\bar q^2) \bar p_1
         - \fraction13 e^{\sqrt2\bar q^1}\cosh(\sqrt6\bar q^2) \bar p_2 \ .
\ee
We can then define an extended Hamiltonian by
\be\label{eq:extham}
     \Hscr = \fraction12 (-\bar p_0{}^2 + \bar p_1{}^2 + \bar p_2{}^2)
        + 2e^{\sqrt2\bar q^1} \cosh\left(\sqrt6\,\bar q^2\right) = 0 \ ,
\ee
where the new momentum is given by $\bar p_0 = \sqrt{2E- \bar p_3{}^2}$. The
next step is to reparametrize the time leading to the Jacobi time gauge
Hamiltonian
\be\label{eq:jacobiham1}
     \Hscr_\Nscr = \fraction12 \Nscr
            (-\bar p_0{}^2 + \bar p_1{}^2 + \bar p_2{}^2) \ ,
\ee
\be\label{eq:jacobiham2}
   \Nscr^{-1} = 2V = 4e^{\sqrt2\bar q^1}\cosh\left(\sqrt6\,\bar q^2\right)\ .
\ee
Although the time reparametrized system is equivalent to the original the
system, old invariants are not necessarily invariants of the new system.
Linear invariants remain invariant after time reparametrization and the new
Hamiltonian is itself a quadratic invariant. However, the cubic
invariant has no immediately obvious counterpart in the new system. A
prescription to transform invariants to the new time gauge is discussed by
Rosquist and Pucacco \cite{rp:inv}. According to that reference the new
invariant is given by $J = \tilde I_3 + R \Hscr_\Nscr$ if we can find a phase
space function $R$ which satisfies the equation
\be\label{eq:rcond}
     \{R, \Hscr\} = \{\tilde I_3, \Nscr^{-1}\}
                                        + \Hscr_\Nscr \{R, \Nscr^{-1}\} \ .
\ee
Straightforward computation shows that
\be
      \{\tilde I_3, \Nscr^{-1}\} = -\fraction{2\sqrt2}{3} e^{\sqrt2x}
     \left[ 2\bar p_x \bar p_y \cosh(\sqrt6y)
            + \sqrt3 \left( \bar p_x{}^2 - \bar p_y{}^2 \right)
             \sinh(\sqrt6y) \right] \ .
\ee
It turns out that the function $R$ can be identified with the linear part
of the the cubic invariant in the old time gauge. More precisely we
have the relations $\{\tilde I_3, \Nscr^{-1}\} = \{R, \Hscr\}$ and $\{R,
\Nscr\} =0$ where $R = -2 \tilde I_3^\ell$ and $\tilde I_3^\ell$ stands for
the linear part of $\tilde I_3$. This shows that $R$ satisfies equation
(\ref{eq:rcond}). Therefore the cubic invariant can be expressed in
the Jacobi time gauge as
\be
     J = - \fraction1{18} \bar p_2{}^3 + \fraction16 \bar p_1{}^2 \bar p_2
         + \fraction1{12} \left[ - \sqrt3 \tanh(\sqrt6 \bar q^2) \bar p_1
                                  + \bar p_2 \right]
          (-\bar p_0{}^2 + \bar p_1{}^2 + \bar p_2{}^2) \ .
\ee
For the purposes of this discussion the important thing to notice about this
expression is that it is homogeneous in the momenta. This fact will be used
to motivate a crucial step in the derivation of the tensorial Lax pair
equation.

\section{The tensorial Lax pair equation}
The basic assumption used to derive the covariant Lax pair equation is that we
have a system which can be described by a purely kinetic Hamiltonian
\be\label{eq:geom_ham}
     H = \fraction12 g^{\alpha\beta}(q) p_\alpha p_\beta \ .
\ee
The equations of motion can then be interpreted as geodesic motion on a
Riemannian or pseudo-Riemannian configuration space with the metric
\be\label{eq:jacobi_metric}
     ds^2 = g_{\alpha\beta}dq^\alpha dq^\beta \ .
\ee
As noted above any Hamiltonian which is quadratic in the momenta can be
transformed to the geometric form (\ref{eq:geom_ham}). Our aim is to obtain a
tensorial form of the Lax equation which is covariant with respect to the
geometry defined by the metric (\ref{eq:jacobi_metric}). Since the right hand
side of (\ref{eq:lax}) involves matrix multiplication we must define the
matrices $\bfL$ and $\bfA$ in such a way that matrix multiplication
corresponds directly to contraction of indices. This can be achieved if the
components of $\bfL$ and $\bfA$ correspond to mixed pairs of indices, {\em
i.e.\/} one covariant and one contravariant index. Choosing the row index to
be contravariant and the column index to be covariant we can then write
\be
    \bfL = (L^\alpha{}_\beta) \ ,\qquad
    \bfA = (A^\alpha{}_\beta) \ .
\ee
At this point we use some properties of the Toda lattice. First, in the
sequence of invariants $I_k= k^{-1} \Tr \bfL^k$, the first one is linear and
homogeneous while the second one is equal to the Hamiltonian itself, $I_2 =
\frac12 \Tr L^2 = H$. Moreover, for geometric Hamiltonians (\ref{eq:geom_ham})
we expect the invariants to be homogeneous in the momenta as in the example of
the Toda lattice. These facts suggest that the elements of $\bfL$ itself
should be homogeneous first order polynomials in the momenta. With this
assumption we have
\be\label{eq:lax_tensor}
     L^\alpha{}_\beta = L^\alpha{}_\beta{}^\gamma p_\gamma \ ,
\ee
where $L^\alpha{}_\beta{}^\gamma$ is some third rank geometric object which
we shall later be able to interpret as a tensor. Taking the time
derivative of (\ref{eq:lax_tensor}) we obtain
\be\label{eq:lhs}
    \dot L^\alpha{}_\beta = \left(
             L^\alpha{}_\beta{}^{(\mu}{}_{,\gamma} \, g^{\nu)\gamma}
           + L^\alpha{}_\beta{}^\gamma \Gamma^{(\mu}{}_\gamma{}^{\nu)}
                                    \right) p_\mu p_\nu \ ,
\ee
where $\Gamma^\alpha{}_{\beta\gamma}$ is the connection of the metric
(\ref{eq:jacobi_metric}). Thus the left hand side of the Lax equation
(\ref{eq:lax}) is quadratic and homogeneous in the momenta. Therefore the
right hand side must also be quadratic and homogeneous. It then becomes
natural to assume that the second Lax matrix $\bfA$ is also a linear and
homogenous function of the momenta. We then have
\be
    A^\alpha{}_\beta = A^\alpha{}_\beta{}^\gamma p_\gamma \ ,
\ee
where $A^\alpha{}_\beta{}^\gamma$ is some geometric object the
nature of which will be specified below. The right hand side of the Lax
equation can now be written
\be\label{eq:rhs}
    ([\bfL,\bfA])^\alpha{}_\beta = \left(
       L^\alpha{}_\gamma{}^\mu A^\gamma{}_\beta{}^\nu
     - A^\alpha{}_\gamma{}^\mu L^\gamma{}_\beta{}^\nu  \right)
       p_\mu p_\nu \ .
\ee

Using the above definitions it follows that the Lax equation (\ref{eq:lax})
can be written as
\be\label{eq:lax2}
    (\dbL - [\bfL,\bfA])^\alpha{}_\beta
     = T^\alpha{}_\beta{}^{\mu\nu} p_\mu p_\nu = 0 \ ,
\ee
where the object $T^\alpha{}_\beta{}  ^{\mu\nu}$ is a function on the
configuration space which can be read off from (\ref{eq:lhs}) and
(\ref{eq:rhs}). Remarkably, $L^\alpha{}_\beta{}^\gamma$ and
$T^\alpha{}_\beta{}^{\mu\nu}$ can be interpreted as tensorial objects related
by the equation
\be
     T^\alpha{}_\beta{}^{\mu\nu} = L^\alpha{}_\beta{}^{(\mu;\nu)} \ ,
\ee
if we make the identification
\be\label{eq:id}
    A^\alpha{}_\beta{}^\gamma = \Gamma^\alpha{}_\beta{}^\gamma \ .
\ee
The result follows directly from the formula for the covariant derivative of
$L^\alpha{}_\beta{}^\gamma$. If the Lax equation is to be satisfied
identically for arbitrary initial data the coefficients of the quadratic
momentum terms in (\ref{eq:lax2}) must vanish separately. This leads to the
equation
\be
      L_{\alpha\beta(\gamma;\delta)} = 0 \ .
\ee

Actually, the above identification is not the most general ansatz which leads
to a tensorial form of the Lax equation. A more general ansatz is to assume
that
\be
      A^\alpha{}_\beta{}^\gamma = \Gamma^\alpha{}_\beta{}^\gamma
                                  + B^\alpha{}_\beta{}^\gamma \ ,
\ee
where $B^\alpha{}_\beta{}^\gamma$ are the
components of a tensor. This leads us to the main result of this
contribution, namely that with this more general ansatz the Lax equation can
be written in the tensorial form
\be\label{eq:lax_tensor_equation}
   L_{\alpha\beta(\gamma;\delta)} =
       L_{\alpha\mu(\gamma} B^\mu{}_{|\beta|\delta)}
     - B_{\alpha\mu(\gamma} L^\mu{}_{|\beta|\delta)} \ ,
\ee
{}From the form of the Toda lattice system one can guess that the antisymmetric
case $B_{\alpha \beta \gamma} = B_{[\alpha \beta] \gamma}$ is of particular
interest. The above equation can then be split into symmetric and
antisymmetric parts by decomposing the Lax tensor itself in symmetric and
antisymmetric parts, $L_{\alpha\beta\gamma} = S_{\alpha\beta\gamma} +
R_{\alpha\beta\gamma}$ where $S_{\alpha\beta\gamma} =
S_{(\alpha\beta)\gamma}$ and $R_{\alpha\beta\gamma} =
R_{[\alpha\beta]\gamma}$. The antisymmetric part then satisfies the
sourceless equation, $R_{\alpha\beta(\gamma;\delta)} = 0$ while the symmetric
Lax tensor equation can be written
\be
     S_{\alpha\beta}{}^{(\gamma;\delta)} =
      -2 S_{(\alpha}{}^{\mu (\gamma} B_{\beta) \mu}{}^{\delta)} \ .
\ee
The Lax pair equation (\ref{eq:lax_tensor_equation}) is in fact a
generalization of the Killing vector equation. Indeed any Lax pair tensor
$L^\alpha{}_\beta{}^\gamma$ with nonvanishing trace on the first two indices
gives rise to the Killing vector $\xi^\alpha = L^\beta{}_\beta{}^\alpha$. Lax
pair tensors $L^\alpha{}_\beta{}^\gamma$ are also generalizations of third
rank Killing-Yano tensors (for a discussion of Killing-Yano tensors in a
general relativistic context see Dietz and R\"udiger \cite{dr:killing-yano}).
This follows immediately from the observation that the Killing-Yano equations
are identical with the sourceless Lax pair equation. Killing-Yano tensors,
however, are by definition 3-forms. Lax pair tensors, on the other hand, have
no {\it a priori\/} symmetry restrictions.

At this point an obvious first problem is to analyze known integrable systems
like {\em e.g.\/} the Toda lattice \cite{toda:book} and the Calogero models
\cite{calogero:inv1} to see if they fit in this framework. The above
discussion of the Toda system makes it very reasonable to believe that at
least the Toda lattice can be given a tensorial Lax pair formulation. A
preliminary investigation has indicated that $B^\alpha{}_\beta{}^\gamma
\neq0$ for the Toda system.

A possible application in general relativity is to look for physically
reasonable spacetimes which admit a Lax pair tensor, for example models of
gravitational collapse. The existence of a Lax pair would imply integrability
of the geodesic equations and thus provide a way to analyze spacetime
singularities analytically. It is an open problem if there exists nontrivial
integrable solutions to the Einstein equations having the Lax pair property.

It is clear from the results of this contribution that one can
define nontrivial integrable spacetimes which are not necessarily solutions
of the Einstein equations. An example is provided by the Hamiltonian
(\ref{eq:jacobiham1}) and its associated metric
\be
     ds^2 = 4e^{\sqrt2\bar q^1}\cosh\left(\sqrt6\,\bar q^2\right)
               \left[ -(dq^0)^2 + (dq^1)^2 + (dq^2)^2 \right] \ .
\ee
This defines a 3-dimensional integrable spacetime having one linear, one
quadratic and one cubic invariant. One can easily extend this construction to
4-dimensional spacetimes for example by considering a four particle Toda
lattice.

\end{document}